\documentstyle[12pt]{article}

\def\mif{{\hspace{0.5cm} \rm if} \hspace{0.5cm}}
\def\mwith{{\hspace{0.5cm} \rm with} \hspace{0.5cm}}

\def\mpunkt{{\hspace{0.5cm} \rm .}}
\def\mkomma{{\hspace{0.5cm} \rm ,} \hspace{0.5cm}}

\begin{document}
\hspace{9cm}Preprint HU Berlin ITP--330\vspace{1.5cm}\\
\noindent
{\Large Entropy and Long range correlations in literary English}
\vspace{0.5cm}\\
Werner Ebeling, Thorsten P\"oschel
\vspace{0.5cm}\\
{\small Institut f\"ur Theoretische Physik, Humboldt--Universit\"at zu Berlin,
Invalidenstra\ss e 42, \\ D-10115 Berlin}\\
\date{{\small\today}}
\vspace{0.5cm}\\
{\bf
Recently long range correlations were detected in nucleotide sequences and in
human writings by
several authors. We undertake here a systematic investigation of two books,
Moby Dick by H.~Melville and Grimm's tales, with respect to the existence of
long
range correlations. The analysis is based on the calculation of entropy like
quantities as the mutual information for pairs of letters and the
entropy, the mean uncertainty, per letter.
We further estimate the number of different subwords of a given
length $n$. Filtering out the contributions due to the
effects of the finite length of the texts, we
find correlations ranging to a few hundred letters. Scaling laws for the
mutual information (decay with a power law), for the entropy per
letter (decay with the inverse square root of $n$) and for the
word numbers (stretched exponential growth with $n$
and with a power law of the text length) were found.}
\renewcommand{\topfraction}{1.0}
\vspace{1cm}\\
{}From a formal point of view a book may be considered as a linear string
of letters. In this respect there exists a similarity to other
linear structures~\cite{ebeling_engel_feistel}. The main
information carriers in living systems
are sequences of amino acids and/or nucleotides, other examples are pieces
of music recorded on
tapes or on paper, computer programs written on disks or tapes etc. By
using the methods of symbolic dynamics any trajectory of a dynamic
system (deterministic or stochastic) may be mapped to a string of letters
on a certain alphabet. Usually the strings generated
by dynamical systems show only short range correlations, except under
critical conditions where, in analogy to equilibrium phase transitions
\cite{stanley},
correlations on all scales may be observed~\cite{ebeling_nicolis_92}.
Recently
long range correlations were detected in DNA sequences~\cite{peng} and in human
writings~\cite{schenkel_92}. The
intrinsic difficulties connected with the analysis of long range correlations
in DNA led to a controversial discussion about the authentic character of
long range structures in DNA~\cite{li_kaneko}.
\par
Naturally the question arises whether long range correlations may be
found in other information carrying strings too. This work is
devoted to the investigation of long range correlations in
texts. We use the methods of entropy analysis, which were first applied to
texts by Claude Shannon in 1951 \cite{shannon}. For several reasons we
expect the existence of long range structures
in these sequences. Since a book is written in a unique style and according
to a general plan of the author, we expect correlations which are
ranging from the beginning of a text up to the end~\cite{epa}.
Already Shannon wrote: ``From this analysis it appears that, in ordinary
literary English, the long range statistical effects (up to 100 letters)
reduce the entropy\dots''.
\par
Another strong argument for long correlations
is based on the combinatorial explosion. Uncorrelated
sequences generated on an alphabet of $\lambda$ letters have a manifold of
$N(n)=\lambda^n=\exp\left[ n\cdot\ln(\lambda)\right]$ different subwords
of length $n$. A subword (block) is here any combination of letters included
the space, punctuation marks and numbers. For $n >100$ the number
$N(n)$ is extremely large. In other words we need very sharp restrictions
to select a meaningful subset. Long range correlations provide
such a strong selection criterion. Hence we we must expect that only a very
small subset $N^*(n)$ of the possible words appears
in a text. Bounds of this kind are given by the rules of writing texts,
i.e. by the rules of syntax as well as by semantic
relations. These rules do not allow for
an arbitrary concatenation of letters to words and of words to sentences but
lead to a limitation of the growth of the number of allowed letter combinations
with $n$. The
problem we address here is, whether the function $N^*(n)$ follows a
simple scaling law.
\par
In earlier papers the conjecture has been made that the number of allowed
subwords scales according to a stretched exponential
law~\cite{ebeling_nicolis_92}\cite{hilberg_90}
\begin{equation}
N^{*}(n)\sim \exp[cn^{\alpha}]  \mwith \alpha \le 1/2 \mkomma c=const.
\label{scaling_rule}
\end{equation}
The reduction due to the scaling rule
(\ref{scaling_rule}) reduces the number of the allowed subwords drastically
($N^*(n) \ll \lambda^n$) for large $n$.
In order to describe a given string of length $L$ using an alphabet of
$\lambda$
letters we introduce the following
notations~\cite{ebeling_nicolis_92}: Let
$A_1 A_2 \dots A_n \nonumber$
be the letters of a given substring of length $n\le L$. Let further
$p^{(n)}(A_1\dots A_n)$
be the probability for this substring (block).
A special case is the probability to find a pair with $(n-2)$
arbitrary letters in between $p^{(n)}(A_1,A_n)$.
Then we may introduce the mutual information for two letters in
distance n (also called
transinformation)~\cite{herzel}\cite{li}:
\begin{equation}
I(n)=\sum_{A_i A_j} p^{(n)}(A_i,A_j)\log
\left[ \frac{p^{(n)}(A_i,A_j)}{p^{(1)}(A_i)\cdot p^{(1)}(A_j)}\right] \mpunkt
\end{equation}
The mutual information is a special measure for correlations
which is closely related to the autocorrelation
function~\cite{peng}\cite{herzel}--\cite{nicolis_katsikas}.
Further we define the entropy per block of length $n$~\cite{grassberger_90}:
\begin{equation}
H_n=-\sum p^{(n)}(A_1 \dots A_n) \log p^{(n)}(A_1 \dots A_n) \mpunkt
\end{equation}
The block entropy is related to the mean number of
words~\cite{ebeling_nicolis_92} by
\begin{equation}
N^{*}(n)\sim \lambda^{H_n} \mpunkt
\label{mcmillanlaw}
\end{equation}
As shown already by Shannon, the entropy per letter of blocks of length n
$H_n/n$ is an important
quantity expressing the structure of sequences.
In~\cite{ebeling_nicolis_92} we assumed the following scaling
behaviour for a definite class of strings at large $n$-values
\begin{equation}
\begin{array}{l}
H_n/n=h + g\cdot n^{\mu_0-1} + e/n \\
0\le \mu_0 <l  \mpunkt
\end{array}
\label{ebeling_scaling}
\end{equation}
Here $h$, the limit of the mean uncertainty, is called the entropy of the
source. This quantity is positive for stochastic
as well as for chaotic processes, $g$ and $e$ are constants; if
$h,e>0$ and $g=0$ the correlations in the string
are short range corresponding to a Markov process with a finite
memory~\cite{grassberger_90}. For periodic strings one finds $h=g=0$, $e>0$.
The existence of a long range order
in strings may be characterized by the condition $g>0$ describing
a slowly decaying contribution to the asymptotics of the entropy per
letter for large $n$. Of special interest for the
further consideration of texts is the case $h=0$, $g>0$ corresponding to
a power law tail of the entropy decaying slower than $1/n$.  A
working hypothesis developed earlier~\cite{ebeling_nicolis_92} is, that
this is the typical behaviour for texts. In other words long texts
are strings on the borderline between periodicity and chaos, showing long
range correlations.
\par
The  mutual information (transinformation) is not a monotonic function
of $n$. We define long range effects by power law tails of the
averaged mutual information $I(n)$. Here the averaging is carried out
over a window comprising several of the typical oscillations
(fluctuations). Several authors have demonstrated that
DNA-sequences show a slowly decaying fluctuations at large scales
{}~\cite{herzel}\cite{li}. As mentioned already, for DNA some
evidence for the existence of long range correlations was
found~\cite{peng}--\cite{li_kaneko}
\cite{herzel}--\cite{nicolis_katsikas}
\cite{herzel_schmitt_ebeling_csf}\cite{voss}.
\par
We will apply the methods of entropy analysis to literary English
represented by the books: ``Moby Dick'' by Melville ($L\approx 1,170,200$)
and Grimm's Tales ($L\approx 1,435,800$).
Let us just mention that pieces of music may
be treated in a similar way~\cite{epa}.
For simplification we use an
alphabet consisting of 32 symbols: the small
letters {\it a b c d e f $\dots$ x y z} \hspace{0.2cm} the marks {\it , . ( )
\# } and
the space; {\it \#} stands for
any number. In order to get a better statistics we have used for the entropy
calculations
also a restricted alphabet consisting of only 3 letters 0, M, L.
The letter 0 codes here for vowels, the letter M stands for consonants and the
letter  L stands  for spaces and marks.
\par
The calculation of the mutual
information requires counting frequencies of pairs of letters at distance $n$.
Since the number of different pairs is $32^2=1024$ we have for our
books a good statistics. The function $I(n)$ is a measure for the
correlations of letters in the distance $n$. Every peak at $n$
corresponds to a positive correlation.
\par
In Fig.~\ref{mutual_moby_grimm} we show
the mutual information
calculated for Moby Dick and for Grimm's Tales
($\lambda=32$). The results show well expressed correlations in
the range $n=1\dots 25$ which are followed by a long slowly decaying tail. The
obtained values for the transinformation $I(k)$ become meaningless
if they are smaller than the level of the fluctuations which
are due to the finite length $L$ of
the text.  According to Herzel~et.~al.~\cite{herzel}
\cite{herzel_schmitt_ebeling_csf} the level of these fluctuations is
\begin{equation}
\delta I(k)=\frac{\lambda^2-2\cdot\lambda}{2\cdot \ln(\lambda)\cdot L} \mpunkt
\end{equation}
For our rather long texts with $L>10^6$ the fluctuation level
has a value of about $10^{-4}$. The smoothed values for the mutual
information for the range $n=25\dots 1000$
may be fitted by the scaling law $I(k)=c_1\cdot n^{-0.37}+c_2$ with
$c_1=1.5\cdot 10^{-4}$, $c_2=1.1\cdot 10^{-4}$. The
constant $c_2$ corresponds here to the level of fluctuations.
Our results prove that long texts show pair correlations
which decay, at least up to distances of several
hundred letters, according to a power law. However due to the greater
uniformity of texts these correlation
tails are not as strong as observed
for DNA sequences~\cite{peng}\cite{li}\cite{herzel_schmitt_ebeling_csf}.
\par
For the calculation of entropies we must count the frequencies of subwords,
where a subword of length n is defined as any combination of $n$ letters.
The result of counting the words consisting of $n=4,~9,~16,~25$ letters
in Grimm's Tales is shown in Fig.~\ref{rank_grimm} in a
rank ordered representation. The structure of the rank ordered
distributions is for both texts rather similar, however the list of words is
of course quite different. For example
among the most frequent subwords of length $n=25$ are in the case of Moby
Dick ``{\it \_greenland\_or\_right\_whale}'', and ``{\it
\_species\_of\_the\_leviathan}''. For
Grimm's Tales rather frequent subwords are e.g. ``{\it
\_if\_i\_could\_but\_shudder.\_}'' and ``{\it princess,\_open\_the\_door\_f}''.
\par
Let us still mention that the form of the subword distributions is
distinctly not Zipf--like, it does not follow a power law. In the opposite,
with
increasing $n$ there is a tendency to form a Fermi--like plateau~\cite
{epa}. This follows from the theorem of asymptotic equipartition derived by
McMillan and Khinchin.
This theorem tells us that for $n\rightarrow
\infty$ the asymptotic form of the distribution is rectangular,
i.e. the $N^*(n)$ allowed subwords of length $n$ appear with nearly
equal frequency. The effects due to finite $n$ and the effects of
finite length $L$ tend to smooth the edges of the
distribution~\cite{herzel_schmitt_ebeling_csf}.
The importance of length corrections
for estimating the frequencies of subwords was considered by several
authors~\cite{herzel}\cite{grassberger_90}. For a deeper analysis of
this problem we refer to
recent articles~\cite{herzel_schmitt_ebeling_csf}\cite{peng_buldyrev}.
Our method
for the entropy analysis uses an extrapolation of the entropy
to infinite text length~\cite{ebeling_nicolis_92}.
We mention also a quite different approach based on the guess of the
distribution function for infinite text
length~\cite{epa}\cite{schmitt_herzel_ebeling_epl}.
\par
The  probabilities which we need for the calculation of entropies are
unknown and can only  be
estimated from the frequencies $N_i(n)$ of the subwords of length $n$ in a text
of length $L$ containing $N=L+1-\lambda$ subwords.
Introducing the observed subword
frequencies into the entropy definition leads to the observed entropies
\begin{equation}
H^{obs}_n=\log(N)-\frac{1}{N}\sum_iN_i(n)\cdot \log\left(N_i(n)\right) \mpunkt
\end{equation}
This is a random variable with the expectation value
\begin{equation}
H^{exp}_n=\langle H^{obs}_n\rangle=\log(N)-\frac{1}{N}\sum_i\langle N_i(n)\cdot
\log
\left(N_i(n)\right)\rangle \mpunkt
\end{equation}
Assuming a Bernoulli distribution for the letter combinations, the mean values
can be calculated explicitly~\cite{herzel}\cite{herzel_schmitt_ebeling_csf}.
The result is
\begin{equation}
H^{exp}_n= \left \{
	\begin{array}{ll}
    	H_n-\frac{N^*(n)}{2 N}  & \mif N^*(n) \ll N \\
	\log (N)-\log (2) \frac{N}{N^*(n)} & \mif N^*(n) \gg N \mpunkt
	\end{array}
	\right.
\label{approximation}
\end{equation}
\par The relation between the effective number of words $N^*(n)$ and the block
entropy $H_n$ is given by eq.~(\ref{mcmillanlaw}).
Hence the expected block entropy may be represented as a function of $\log N$
with one free parameter $H_n$ which is found by fitting the curves.
In this way the block entropies for both books were calculated up to $n=26$.
For small word length we used the approximation (\ref{approximation}) for
$N^*(n)\ll N$ and
for larger $n$ we applied the approximation valid for $N^*(n)\gg N$. In the
intermediate region we applied a smooth Pad\'{e} approximation between both
formulae. In a procedure of successive approximations the entropy
$H_n$ was considered as a free parameter which was fitted in a way that
$H^{exp}_n (\log N)$ came as close as possible to the measured (observed)
entropy values. In practice this method breaks down for $n> 30$ if
$\lambda = 3$ and for $n>25$ if $\lambda = 32$. Longer subwords do
not have a chance to appear several times in the text, what leads to
large statistical errors.
\par
The calculations for $n \le 26$ show that the square root law yields a
reasonable approximation for the
scaling of the entropy per letter with the word length $n$
\begin{equation}
\begin{array}{llllll}
H_n/(n \cdot \log(\lambda)) & \approx & (4.84/ \sqrt{n}) & -(7.57/n) &
\hspace{0.1cm}(\lambda =3) \\
H_n/(n \cdot \log(\lambda)) & \approx & (0.9/ \sqrt{n}) & + (1.7/n) &
\hspace{0.1cm}(\lambda =32)\mpunkt
\end{array}
\end{equation}
Fig.~\ref{moby_entr} shows the fit for the alphabet $\lambda=3$.
The scaling law of the square root type was first found
by Hilberg~\cite{hilberg_90} by fitting Shannons original data. For $n=100$
and $\lambda=32$ our scaling formula gives $H_{100}\approx 10\cdot
\log(\lambda)$ what
is not far from Shannon's estimation $H_{100}\approx 40$ bits.
\par
The number of subwords increases according to a stressed exponential law. For
the law of growth we found the approximation
\begin{eqnarray}
N_n^*\approx & 2^{23.5 \sqrt{n}-35.5} & \hspace{1cm}(\lambda=3)\\
N_n^*\approx & 2^{4.5 \sqrt{n}+8.5} & \hspace{1cm} (\lambda=32)\mpunkt
\end{eqnarray}
We summarize now the results obtained for the two books: The scaling of
the mutual information and the entropy per letter shows in
agreement with earlier work~\cite{ebeling_nicolis_92}
that  long texts  are neither periodic
nor chaotic but somehow in between.
We found correlations in the  range  up
to $10^3$  positions. The existence of such correlations is to be seen
in the statistics of pairs of letters and of blocks (subwords) of letters.
We developed  methods, for the calculation of entropies
from the given samples of limited length. Taking into account length
corrections we calculated block entropies up to $n=26$ and mutual
informations up to distances of a few hundred letters. Based on these
data we formulated a hypothesis about the long range scaling.
For the range $n \gg 100$ the pair correlations
contained in the transinformation of long texts $L>10^6$ decay according
to a power law, however the differences to Bernoulli samples of the
same length are rather small. A reliable estimation of the block entropies
for $n>30$ is still an open question. The results for the entropy
of the two books suggest in agreement with Shannon`s data and Hilberg`s
findings that the mean entropy per letter decays to zero according to
a square root law. As a consequence
the number of different subwords in texts increases with the number of letters
$n$
according to a stretched exponential law.
Our estimations for the growth yield for $n=100$
a total number of about $2^{53}$ different subwords. In spite
of the fact that this number is very large, it is indeed small in
comparison to Bernoulli strings where $2^{108}$ different subwords of
length $n=100$ exist. In this way we observe a very strong
selection among the combinatorial possibilities. Most of the subwords
which would be possible from the combinatorial
point of view are actually forbidden and do not appear in real texts.
We investigated also how the number of genuine english words $N(L)$
(formally defined here as sequences of letters between
spaces and/or marks) increases with the
length $L$ of a text. For Grimm's Tales we found the scaling law
$N(L)=22.8 \cdot L^{0.46}$. In other words with increasing length always new
words are
introduced, no saturation with text length could be observed.
\par
More empirical data on long texts and further studies of the statistical
effects due to finite length of the samples are needed in order to reach
a more definite conclusion about the scaling properties.
\par
The  authors  thank  K.~Albrecht, J.~Freund, H.~Herzel, G.~Nicolis and
A.~Schmitt for fruitful discussions and collaboration on several
tasks of the methods leading to the results presented here.

\newpage
\begin{figure}[ht]
\caption{\it The mutual information calculated for  Melville's Moby Dick
and for Grimm's Tales ($\lambda=32$, $n< 25$).}
\label{mutual_moby_grimm}
\end{figure}
\begin{figure}[ht]
\caption{\it The observed rank ordered distribution of words of length
$n=4, 9, 16, 25$ for Grimm's Tales.}
\label{rank_grimm}
\end{figure}
\begin{figure}[ht]
\caption{The scaling behaviour of the block entropy $H_n$ with the square root
of
the word length $n$ for Moby Dick encoded by the alphabet $\lambda=3$.}
\label{moby_entr}
\end{figure}
\end{document}